\input epsf

\magnification\magstephalf
\overfullrule 0pt
\def\gsim{\raise.3ex\hbox{$\;>$\kern-.75em\lower1ex\hbox{$\sim$}$\;$}}

\font\rfont=cmr10 at 10 true pt
\def\ref#1{$^{\hbox{\rfont {[#1]}}}$}

  %%Fonts

\font\fourteenbf=cmbx12 scaled\magstep1

\font\tenbfit=cmbxti10
\font\sevenbfit=cmbxti10 at 7pt
\font\fivebfit=cmbxti10 at 5pt
\newfam\bfitfam 
\textfont\bfitfam=\tenbfit  \scriptfont\bfitfam=\sevenbfit
\scriptscriptfont\bfitfam=\fivebfit

\font\eightit=cmti8

%%%family 8%%%
\font\tenbfit=cmbxti10
\font\sevenbfit=cmbxti10 at 7pt
\font\fivebfit=cmbxti10 at 5pt
\newfam\bfitfam 
\textfont\bfitfam=\tenbfit  \scriptfont\bfitfam=\sevenbfit
\scriptscriptfont\bfitfam=\fivebfit

%%%family 9%%%
\font\tenbit=cmmib10
\newfam\bitfam
\textfont\bitfam=\tenbit%

%%%family 10%%%
\font\tenmbf=cmbx10
\font\sevenmbf=cmbx7
\font\fivembf=cmbx5
\newfam\mbffam
\textfont\mbffam=\tenmbf \scriptfont\mbffam=\sevenmbf
\scriptscriptfont\mbffam=\fivembf

%%%family 11%%%
\font\tenbsy=cmbsy10
\newfam\bsyfam 
\textfont\bsyfam=\tenbsy%

  %%Greek

\def\pmb#1{\setbox0=\hbox{#1}% \kern-.025em\copy0\kern-\wd0
 \kern.05em\copy0\kern-\wd0 \kern-.025em\raise.0433em\box0 }

\def\slash{/\kern-.5em}

  %%Fractions
\def \half {{\textstyle {1 \over 2}}}

 %

 %%FORMATTING

\def\boxit#1{\vbox{\hrule\hbox{\vrule\kern1pt\vbox
{\kern1pt#1\kern1pt}\kern1pt\vrule}\hrule}}

\def\h{\hfill\break}
\parskip=6pt
\parindent=0pt
\hsize=17truecm\hoffset=-5truemm
\vsize=23truecm
\def\footnoterule{\kern-3pt
\hrule width 17truecm \kern 2.6pt}

  %%REFERENCES
%     \defref\label{text}
% generates a number, assigns it to \label, generates an entry.
% To list the refs,  \listrefs
% (Extracted and adapted from harvmac.tex by P Ginsparg)

\catcode`\@=11 % This allows us to modify PLAIN macros.

\def\nolabels{\def\wrlabeL##1{}\def\eqlabeL##1{}\def\reflabeL##1{}}
\def\writelabels{\def\wrlabeL##1{\leavevmode\vadjust{\rlap{\smash%
{\line{{\escapechar=` \hfill\rlap{\sevenrm\hskip.03in\string##1}}}}}}}%
\def\eqlabeL##1{{\escapechar-1\rlap{\sevenrm\hskip.05in\string##1}}}%
\def\reflabeL##1{\noexpand\llap{\noexpand\sevenrm\string\string\string##1}}}
\nolabels
\global\newcount\refno \global\refno=1
\newwrite\rfile
\def\defref{$^{{\hbox{\rfont [\the\refno]}}}$\nref}
\def\nref#1{\xdef#1{\the\refno}\writedef{#1\leftbracket#1}%
\ifnum\refno=1\immediate\openout\rfile=refs.tmp\fi
\global\advance\refno by1\chardef\wfile=\rfile\immediate
\write\rfile{\noexpand\item{#1\ }\reflabeL{#1\hskip.31in}\pctsign}\findarg}
%	horrible hack to sidestep tex \write limitation
\def\findarg#1#{\begingroup\obeylines\newlinechar=`\^^M\pass@rg}
{\obeylines\gdef\pass@rg#1{\writ@line\relax #1^^M\hbox{}^^M}%
\gdef\writ@line#1^^M{\expandafter\toks0\expandafter{\striprel@x #1}%
\edef\next{\the\toks0}\ifx\next\em@rk\let\next=\endgroup\else\ifx\next\empty%
\else\immediate\write\wfile{\the\toks0}\fi\let\next=\writ@line\fi\next\relax}}
\def\striprel@x#1{} \def\em@rk{\hbox{}} 
\def\lref{\begingroup\obeylines\lr@f}
\def\lr@f#1#2{\gdef#1{\defref#1{#2}}\endgroup\unskip}
\newwrite\lfile
{\escapechar-1\xdef\pctsign{\string\%}\xdef\leftbracket{\string\{}
\xdef\rightbracket{\string\}}}

\def\writestop{\def\writestoppt{\immediate\write\lfile{\string\p
ageno%
\the\pageno\string\startrefs\leftbracket\the\refno\rightbracket%
\string\def\string\secsym\leftbracket\secsym\rightbracket%
\string\secno\the\secno\string\meqno\the\meqno}\immediate\closeout\lfile}}
\def\writestoppt{}\def\writedef#1{}
\catcode`\@=12 % at signs are no longer letters
\rightline{M/C-TH 99-16}
\rightline{DAMTP-1999-167}
\vskip 7pt
\centerline{\fourteenbf EXCLUSIVE VECTOR PHOTOPRODUCTION:}
\vskip 3pt
\centerline{\fourteenbf CONFIRMATION OF REGGE THEORY}
\vskip 8pt
\centerline{A Donnachie}
\centerline{Department of Physics, Manchester University}
\vskip 5pt
\centerline{P V Landshoff}
\centerline{DAMTP, Cambridge University$^*$}
\footnote{}{$^*$ email addresses: ad@a3.ph.man.ac.uk, \ pvl@damtp.cam.ac.uk}
\bigskip
{\bf Abstract}
\vskip 1mm
Recent small-$t$ ZEUS data for exclusive 
$\rho$ photoproduction are in excellent
agreement with exchange of the classical soft pomeron with slope 
$\alpha '=0.25$ GeV$^{-2}$. Adding in a flavour-blind hard-pomeron
contribution, whose magnitude is calculated from the data for
exclusive $J/\psi$ photoproduction, gives a good fit also to
the ZEUS data for $\rho$ photoproduction at larger values of $t$, and to
$\phi$ photoproduction.

\vskip 15mm

The ZEUS collaboration has recently suggested\defref\zeusrho{
ZEUS collaboration: J Breitweg et al, Europ Phys Jour  C1 (1998) 81
and hep-ex/9910038
}
that their data for exclusive $\rho$ photoproduction at HERA,
when combined with lower-energy data\defref\lowrho{
D Aston et al, Nuclear Physics B209 (1982) 56
}
lead to
a slope $\alpha '$ for the trajectory of the soft pomeron that differs
significantly from the classical value $\alpha '=0.25$ GeV$^{-2}$. A main 
message of this paper is to disagree with this conclusion; we show
that in fact the classical value is confirmed by the data.

The slope $\alpha '$ should be determined from the data at small $t$, but
the ZEUS measurements extend also to rather large $t$. 
At HERA energy, soft-pomeron
exchange dominates the differential cross-section out to values of $|t|$
of about 0.4 GeV$^2$. Beyond that, some new contribution is needed.
For exclusive $J/\psi$ photoproduction, a new contribution is needed
even at very small $t$. We have shown recently\defref\charm{
A Donnachie and P V Landshoff, hep-ph/9910262}
that introducing a ``hard pomeron''  gives an excellent description of
data not only for $J/\psi$ photoproduction, but also for the charm
structure function $F_2^c$ and the small-$x$ behaviour of the total
structure function $F_2$. A second message in the present paper is
that the introduction of the same hard pomeron
also provides a good description of the large-$t$ $\rho$~photoproduction
data. Applying the model to $\phi$ photoproduction gives a satisfactory 
description of these data too. 
\topinsert
\centerline{\epsfxsize=0.6\hsize\epsfbox{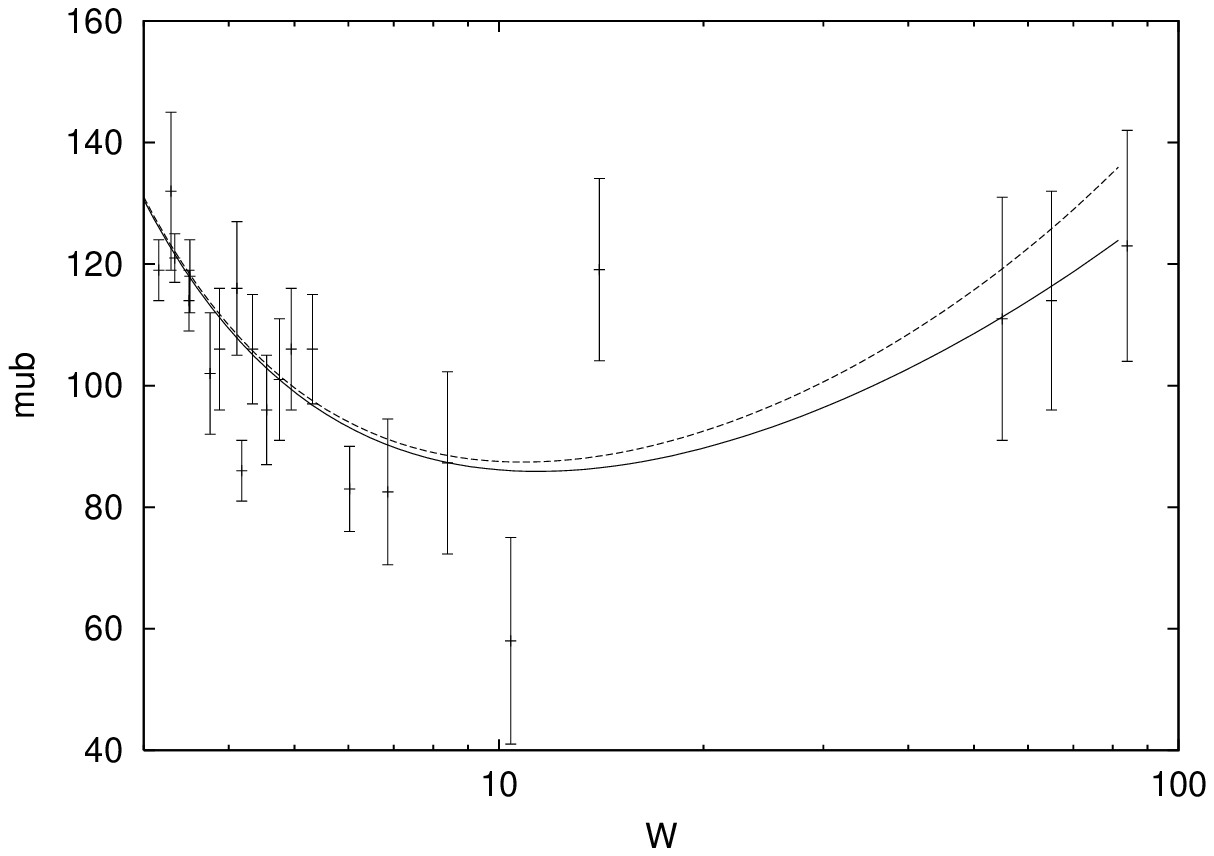}}
Figure 1: Data for the forward differential cross section
for exclusive $\rho ^0$ photoproduction. The solid curve corresponds to
the exchange of the soft pomeron together with $f$ and $a_2$, while the dashed
curve includes also a hard-pomeron contribution.
\endinsert

Consider first small-$t$ $\rho$ photoproduction. In order to extract
the soft-pomeron slope $\alpha '$, it is necessary to consider data
from HERA together with measurements at much lower energy. We have shown 
previously\defref\vector{
A Donnachie and P V Landshoff, Physics Letters B348 (1995) 213
}
that the description of the lower-energy data
needs a significant
contribution from $f$ and $a_2$ exchange; this is missing from
the ZEUS analysis\ref{\zeusrho}. Further, as is apparent from
the data shown in figure 1, the relative normalisation of 
the lower energy experiments is somewhat erratic and it is not correct to 
use just one or two energies for comparison. 
It is necessary to perform a global 
fit to average out the discrepancies.
In our previous analysis\ref{\vector}, we first
assumed that the contribution from soft-pomeron and $f,a_2$ exchanges
to the $\rho ^0p$ total cross section is the same as in the average of 
the $\pi^+p$ and $\pi^-p$ cross sections\defref\sigtot{
A Donnachie and P V Landshoff, Physics Letters B296 (1992) 227
}. 
We then used $\rho$-dominance, with a factor of 0.84 to allow for 
finite-width corrections to $\rho\to e^+e^-$ decay\defref\gounaris{
G Gounaris and J J Sakurai, Physical Review Letters 21 (1968) 244\h
F M Renard, Nuclear Physics B15 (1970) 267
},
to calculate the forward differential cross section for
$\gamma p\to\rho p$.  Figure 1 (solid curve)
shows the resulting cross section at $t = 0$ as a function of energy.

\goodbreak
In order to extend this away from the forward direction, 
we need the two Regge trajectories
$$
\alpha _{P_{1}}(t)=1.08+\alpha '_{P_{1}} t~~~~~~~~~~~~~~~~
\alpha '_{P_{1}}=0.25$$$$
\alpha _R(t)=0.55+\alpha '_R t~~~~~~~~~~~~~~~~ \alpha '_R=0.93
\eqno(1)
$$
We need also to decide the mass scale $s_0$ by which we must divide $s$
before we raise it to the Regge power. There is no theory that determines
this. We adopt the dual-model prescription\defref\veneziano{
G Veneziano, Nuovo Cimento 57A (1968) 190
}
that, for a trajectory of slope $\alpha '$, one should take $s_0=1/\alpha '$. 
It is well-established\defref\elastic{
A Donnachie and P V Landshoff, Nuclear Physics B231 (1983) 189
}
that the trajectories couple to the proton through the Dirac electric 
form factor
$$
F(t)={4m^2-2.79t \over 4m^2-t} \left ({1\over 1-t/0.71}\right )^2
\eqno(2)
$$
but their coupling $G_{\rho}(t)$ to the $\gamma\rho$ vertex is unknown. We find that
a good description of the data is provided by the choice
$$
G_{\rho}(t)={1\over 1-t/0.71}
\eqno(3)
$$
Putting these things together, we have for the soft part of the amplitude
for $\gamma p\to\rho p$
$$
T_{\hbox{\fiverm SOFT}}(s,t)=iF(t)G_{\rho}(t)\Big [A_{P_{1}}
(\alpha '_{P_{1}}s)^{\alpha _{P_{1}}(t)-1}
e^{-\half i\pi(\alpha _{P_{1}}(t)-1)}+
A_{R} (\alpha '_{R}s)^{\alpha _{R}(t)-1}e^{-\half i\pi(\alpha _{R}(t)-1)}
\Big ]
\eqno(4a)
$$
with
$$
A_{P_{1}}=6.0 ~~~~~~~~~~~~~~~~ A_{R}=15.9
\eqno(4b)
$$
The amplitude is normalised such that $d\sigma/dt=|T(s,t)|^2$ in 
$\mu b$ GeV$^{-2}$.
\topinsert
\line{\hfill\epsfxsize=0.4\hsize\epsfbox{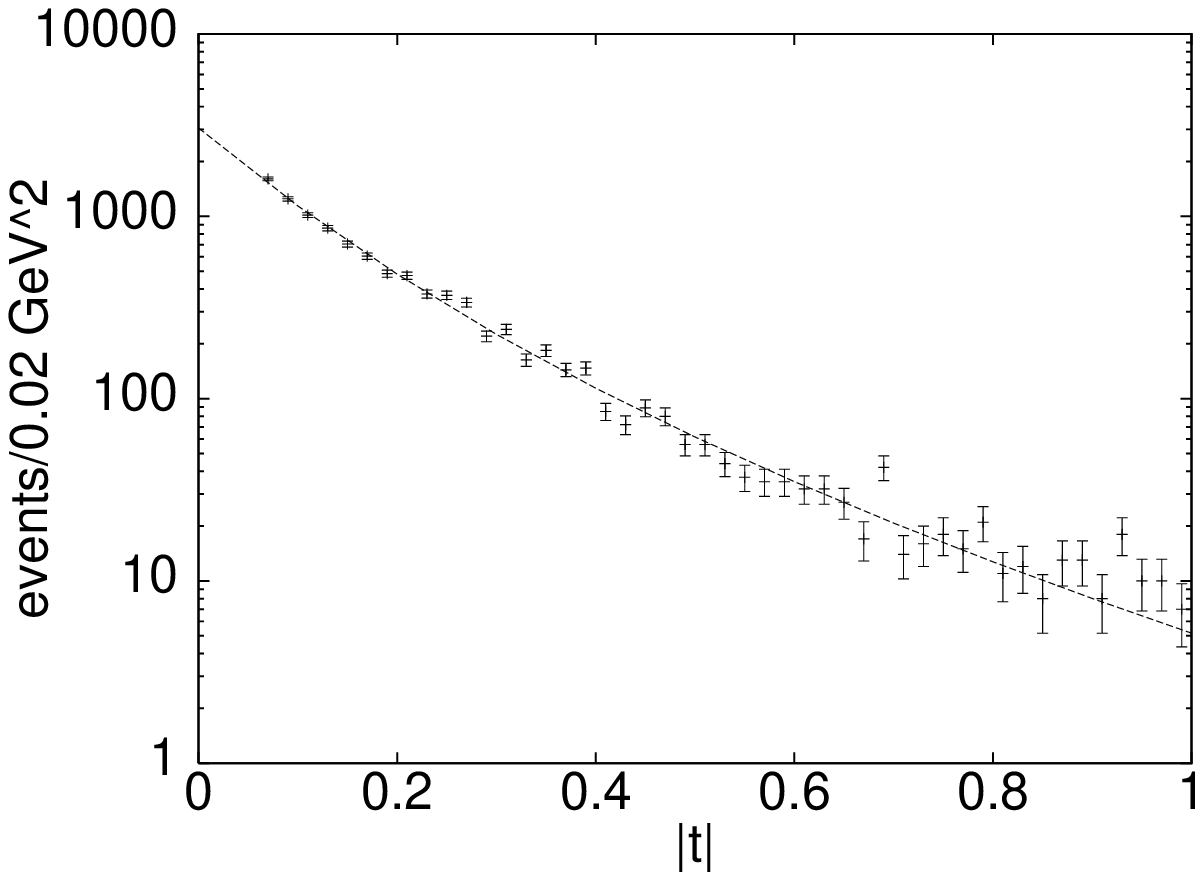}\hfill
\epsfxsize=0.4\hsize\epsfbox{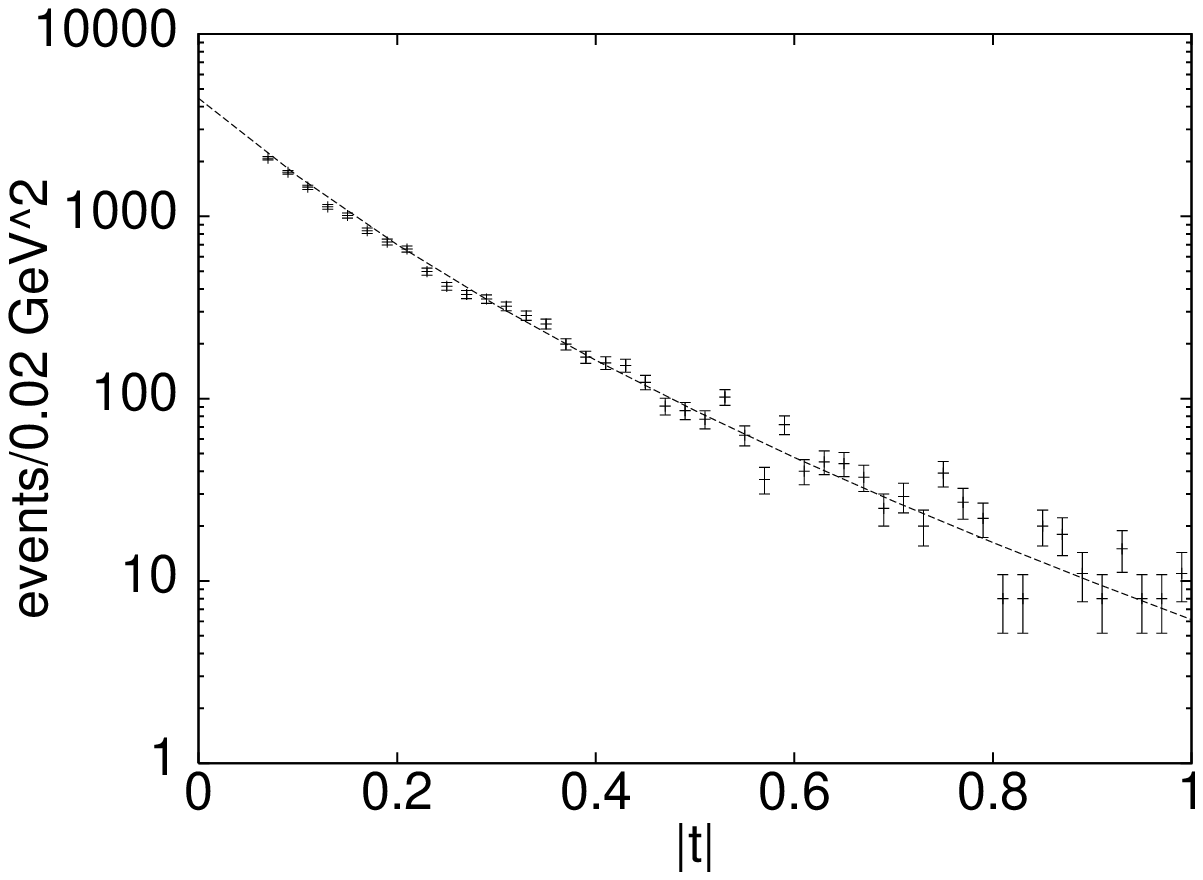}\hfill}
Figure 2: Data\ref{\lowrho}
for $\gamma p\to\rho p$ at $\surd s= 6.86$ and 10.4 GeV,
with Regge fit.
\vskip 10mm
\line{\hfill\epsfxsize=0.6\hsize\epsfbox{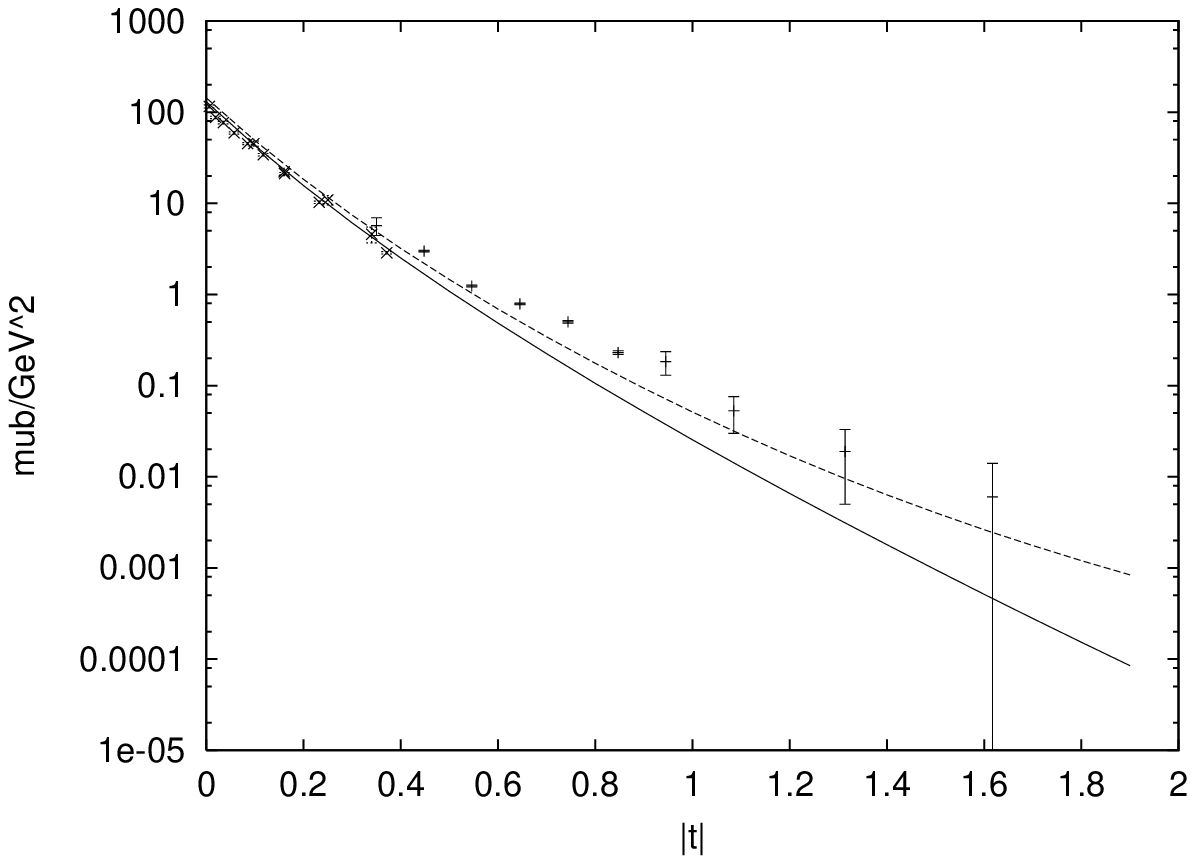}\hfill}
Figure 3: ZEUS data\ref{\zeusrho} for $\gamma p\to\rho p$. The lower-$t$ 
data are at  $\surd s=$71.7~GeV and the higher-$t$ at 94 GeV.
The solid line is the Regge fit with soft exchanges only; the dashed line
includes the hard pomeron (the fits change very little between the two
energies)
\endinsert

Figure 2 shows the differential cross section at $\surd s=$6.86
and 10.4~GeV, together with CERN Omega data\ref{\lowrho}. The data are not 
normalised. The solid line in
figure 3 shows the same fit at $\surd s=$94~GeV, together with ZEUS
data\ref{\zeusrho} (which are normalised). The success of the fit at 
small $t$ is evidence that 
the classical value 0.25 of the soft-pomeron slope $\alpha '_{P_1}$ is 
correct. The dashed lines in figures 1 and 3 include an additional contribution
which we now discuss.

What we have said so far should be uncontroversial. The remainder of this
paper concerns the hard pomeron and so may be less generally accepted, 
though it adds to the already-strong body of evidence in support of the
concept. 
We first introduced\defref\twopom{
A Donnachie and P V Landshoff, Physics Letters B437 (1998) 408 
}
a hard pomeron, with intercept $\alpha _{P_0}$ a little
greater than 1.4, to explain the data for the proton structure function
$F_2(x,Q^2)$ at small $x$. We then observed\ref{\charm} that the ZEUS 
data\defref\zeuscharm{
ZEUS collaboration: A Breitweg et al, hep-ex/9908012
}
for the charm component $F^c_2(x,Q^2)$ of $F_2(x,Q^2)$ seem to confirm
its existence, and tentatively deduced the slope of the trajectory from the
H1 data\defref\hone{
H1 Collaboration, submitted to
the International Europhysics Conference on High Energy Physics HEP99, Tampere,
Finland, July 1999
}
for the differential cross section for the process
$\gamma p\to J/\psi\,p$:
$$
\alpha _{P_{0}}(t)=1.44+\alpha '_{P_{0}} t~~~~~~~~~~~~~~~~
\alpha '_{P_{0}}=0.1
\eqno(5)
$$
We found also that the coupling $G_{J/\psi}(t)$ to the $\gamma J/\psi$ vertex
is rather flatter in $t$ than the coupling $G_{\rho}(t)$ to the
$\gamma\rho$ vertex that we specify in (3), and we took it to be constant.

\topinsert
\line{\hfill\epsfxsize=0.4\hsize\epsfbox{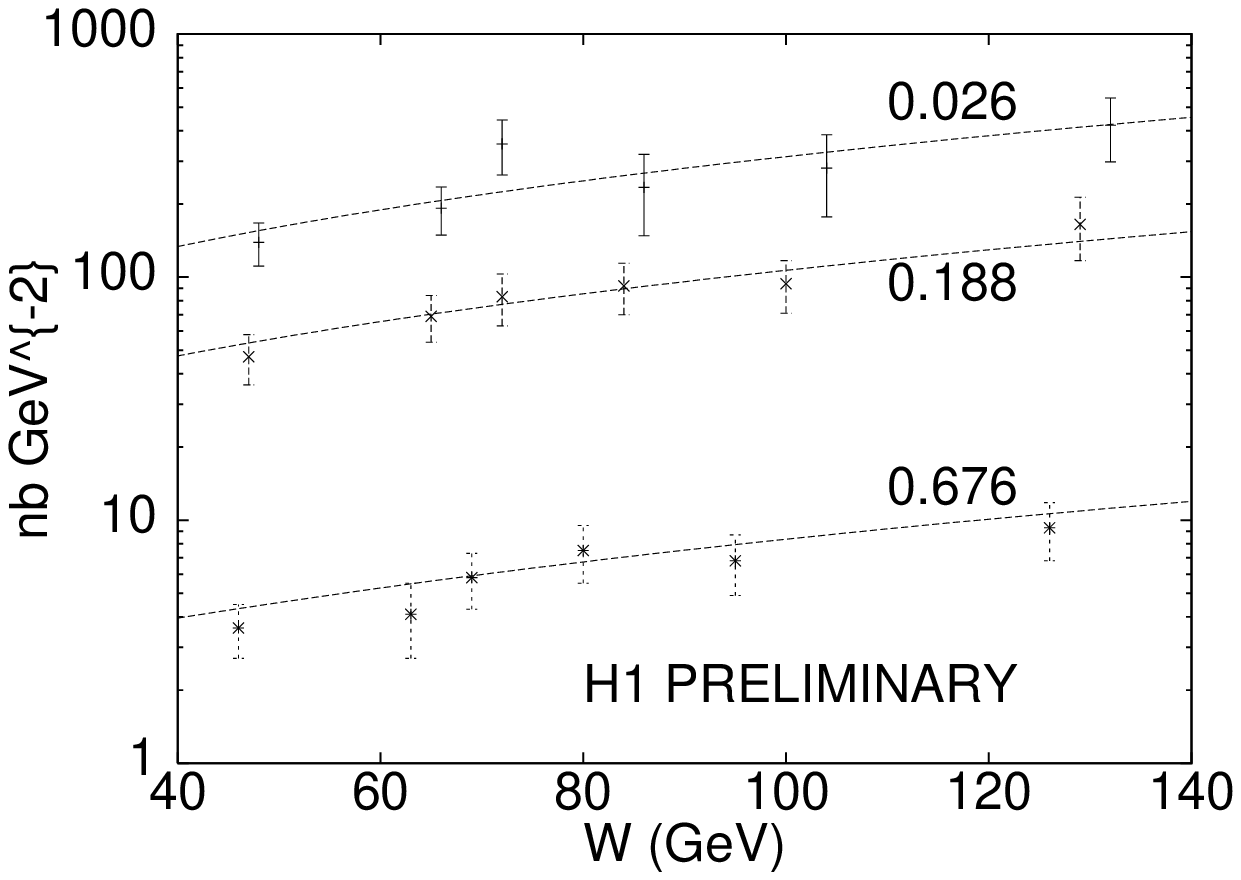}\hfill
\epsfxsize=0.4\hsize\epsfbox{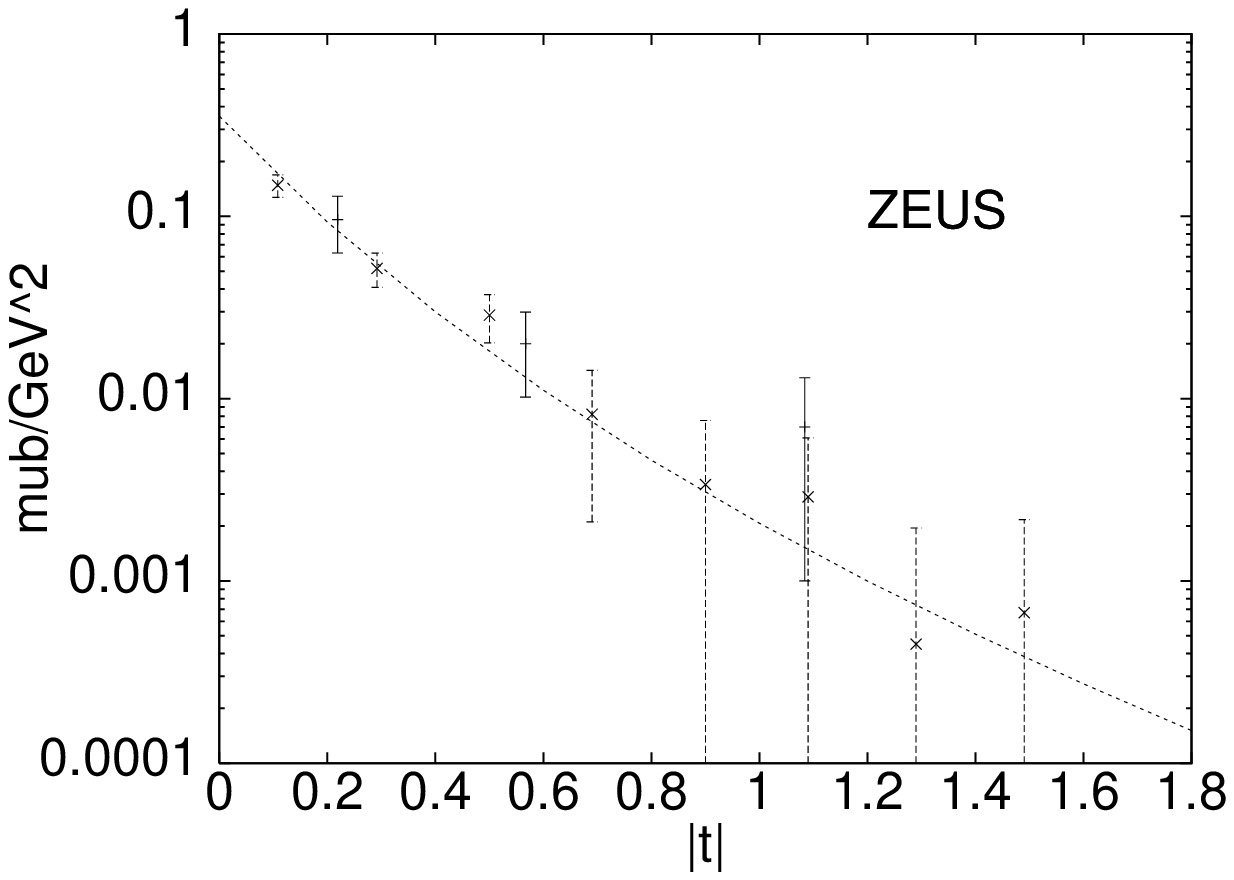}\hfill}
Figure 4: $\gamma p\to J/\psi\,p$: H1 data\ref{\hone}
 at three $t$ values, and ZEUS
data\ref{\zeusrho}
at $\surd s$=94 GeV. The fits include the hard and soft pomeron
contributions
\endinsert

So the hard-pomeron contribution to the amplitude for 
$\gamma p\to J/\psi\,p$ is taken as
$$
iF(t)\Big [A_{P_{0}}
(\alpha '_{P_{0}}s)^{\alpha _{P_{0}}(t)-1}
e^{-\half i\pi(\alpha _{P_{0}}(t)-1)}
\Big ]
\eqno(6)
$$
This differs from what we used in reference [{\vector}] in that we 
divide $s$ by the mass scale $s_0=1/\alpha '_{P_0}$
before raising it to the Regge power. We assume Zweig's rule, so
that the $f,a_2$ trajectory decouples and $A_{R}=0$, although a 
contribution from the soft pomeron is retained. 
Figure 4 shows
the comparison with the H1 data\ref{\hone}
at three values of $t$, and the ZEUS data\ref{\zeusrho}
at $\surd s$=94 GeV, taking
$$
A_{P_0}=0.016~~~~~~~~~~~~~A_{P_1}=0.17
\eqno(7)
$$

Our fits to the data for $F_2$ at small $x$ and for $F_2^c$ suggest that
the coupling of the pomeron to quarks is flavour-blind. So, in order to
relate the strengths of the hard-pomeron couplings in the processes
$\gamma p\to J/\psi\,p$ and $\gamma p\to\rho p$ we need just to include
wave-function effects. Although the hard pomeron couples to photon-induced
reactions, its coupling to purely hadronic processes is extremely 
small\ref{\twopom}. So it seems reasonable to assume that it is the pointlike
component of the photon that is largely responsible, rather than the 
hadron-like component. This in turn implies
that in $\gamma p\to V p$, the strength of the hard-pomeron coupling
depends on the magnitude of the $V$ wave function at the origin and
the relevant quark charges, and
that therefore it is proportional to $\sqrt{\Gamma _{V\to e^+e^-}/m_V}$.
This implies that for $\gamma p\to\rho p$ we should use
$$
A_{P_0}=0.036
\eqno(8)
$$
Adding such a hard-pomeron term to the amplitude gives the dashed curve in 
figure 3.

It might have been thought that a Regge cut, for example from two-pomeron 
exchange, could have been used to explain the $\rho$ data at larger $|t|$ 
as the cut has a less strong t-dependence than the pole. However the cut
has the opposite sign to the pole, so far from enhancing the cross section 
at large $|t|$ it actually reduces it.  

Finally we apply the model to $\phi$ photoproduction. As before we can use 
the flavour-blind nature of the coupling of the hard pomeron to quarks to 
specify its contribution uniquely. This gives
$$
A_{P_0}=0.014
\eqno(9)
$$
Just as for the $\rho$ there are two unknowns in the soft pomeron 
contribution to $\phi$ photoproduction: the magnitude of the coupling of 
the soft pomeron to strange quarks and the mass scale in the $\phi$ form 
factor. We know that Vector Meson Dominance is not a good approximation 
for the $\phi$ and that wave function effects are important \ref{\vector}, 
so the normalisation can only be specified by the data. Naively the mass
in the form factor is simply that of the $\phi$ but higher-mass $s\bar{s}$
states must contribute making the effective mass somewhat larger. In analogy 
with the $\rho$ case we use
$$
G_{\phi} = {{1}\over{1-t/1.5}}
\eqno(10)
$$
choosing 1.5 instead of 0.71 on the grounds that $m_\phi^2 \sim 2m_\rho^2$.
As for the $J/\psi$ we can neglect any contribution from $f,a_2$ exchange.
Fitting the soft pomeron coupling to the data yields
$$
A_{P_1}=1.49
\eqno(11)
$$ 
and the results are shown in figure 5.
\bigskip
\midinsert
\line{\hfill\epsfxsize=0.4\hsize\epsfbox{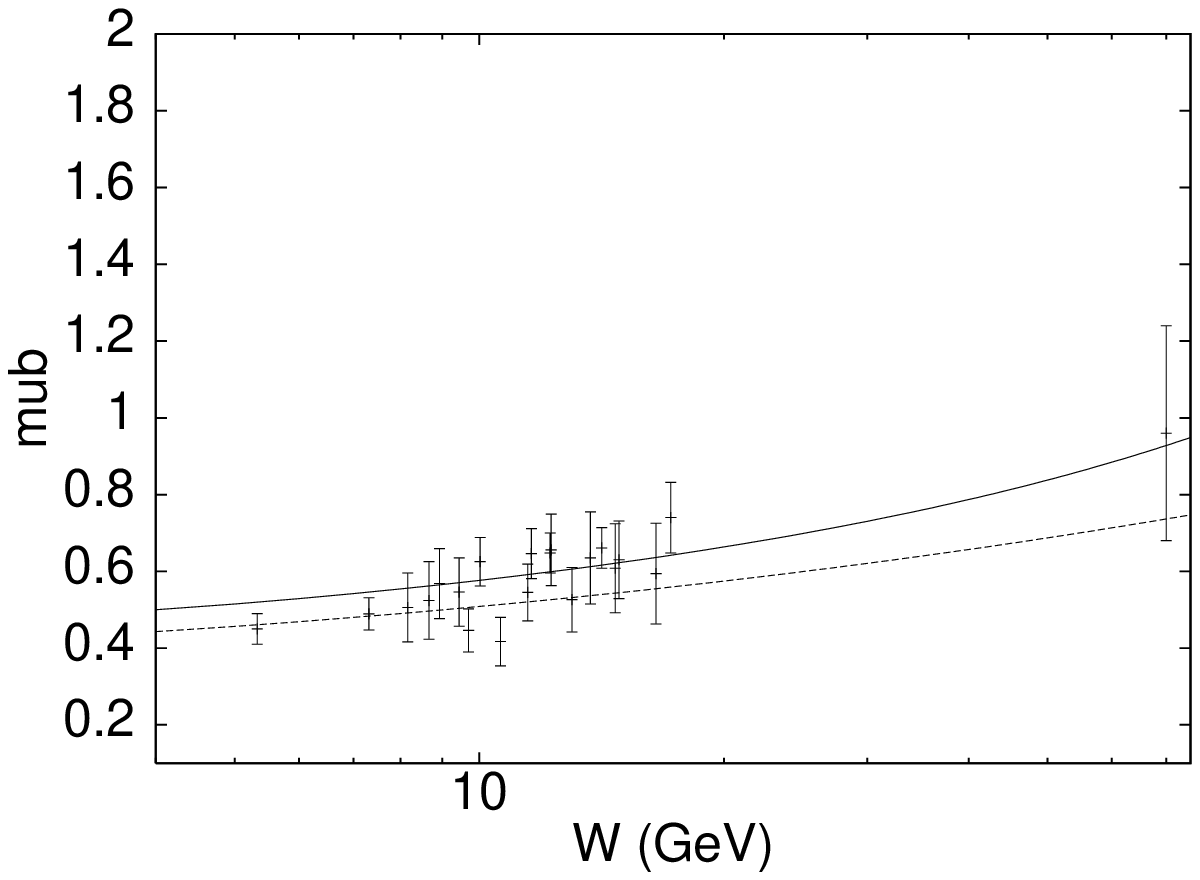}\hfill
\epsfxsize=0.4\hsize\epsfbox{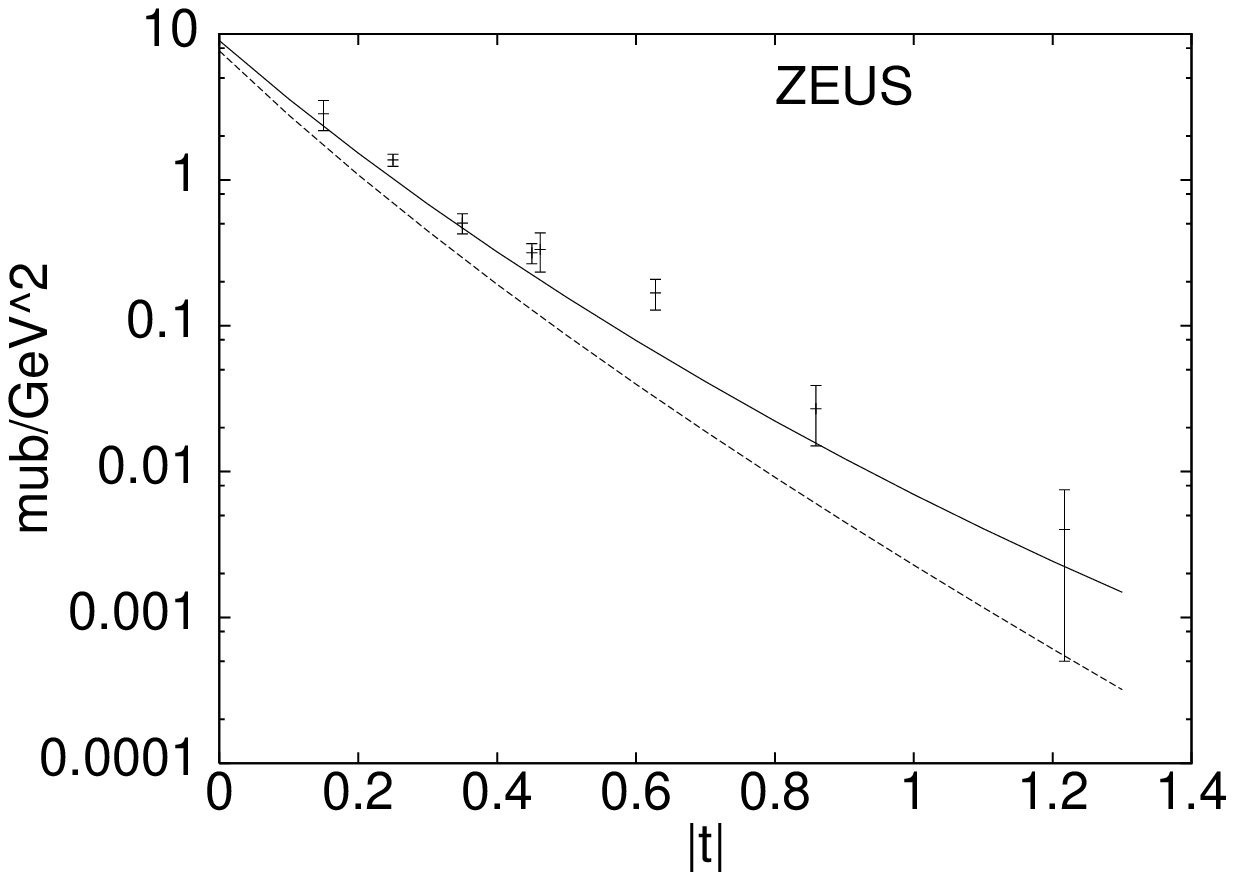}\hfill}
Figure 5: $\gamma p\to \phi\,p$:  Data for the total cross section\defref\bus{
J Busenitz et al, Physical Review D40 (1989) 1
} and the differential cross section\ref{\zeusrho}\defref\zeusphi{
ZEUS collaboration: M Derrick et al, Physics Letters B377 (1996) 259
}
at $\surd s$=94 GeV. 
The dashed lines show the soft pomeron contributions and the solid lines
include also the hard pomeron
\endinsert

\goodbreak
\bigskip{\eightit
This research is supported in part by the EU Programme
``Training and Mobility of Researchers", Networks
``Hadronic Physics with High Energy Electromagnetic Probes"
(contract FMRX-CT96-0008) and
``Quantum Chromodynamics and the Deep Structure of
Elementary Particles'' (contract FMRX-CT98-0194),
and by PPARC}
\goodbreak
\bigskip
\medskip\immediate\closeout\rfile\writestoppt
\baselineskip=10pt{{\bf References}}\bigskip{\frenchspacing%
\parindent=20pt\escapechar=` \input refs.tmp\bigskip}\nonfrenchspacing

\bye